\documentclass[preprint,prb,showpacs,preprintnumbers,amsmath,amssymb]{revtex4}

\usepackage{graphicx}
\usepackage{dcolumn}
\usepackage{bm}

\begin{document}

\preprint{APS/PRB}

\title{Inelastic neutron scattering study of the magnetic fluctuations in Sr$_2$RuO$_4$}

\author{K.~Iida$^1$}
\author{M.~Kofu$^1$}\altaffiliation{Present address: Neutron Science Laboratory, Institute for Solid State Physics, University of Tokyo, Kashiwa, Chiba 277-8581, Japan}
\author{N.~Katayama$^1$}
\author{J.~Lee$^{1,2}$}
\author{R.~Kajimoto$^3$}
\author{Y.~Inamura$^3$}
\author{M.~Nakamura$^3$}
\author{M.~Arai$^3$}
\author{Y.~Yoshida$^4$}
\author{M.~Fujita$^5$}
\author{K.~Yamada$^5$}
\author{S.-H.~Lee$^1$}\email{shlee@virginia.edu}

\affiliation{$^1$Department of Physics, University of Virginia, Charlottesville, Virginia 22904-4714, USA}
\affiliation{$^2$Neutron Scattering Science Division, Oak Ridge National Laboratory, Oak Ridge, Tennessee 37831, USA}
\affiliation{$^3$Materials and Life Science Division, J-PARC Center, Tokai, Ibaraki 319-1195, Japan}
\affiliation{$^4$National Institute of Advanced Industrial Science and Technology, Tsukuba, Ibaraki 305-8565, Japan}
\affiliation{$^5$WPI Research Center, Advanced Institute for Materials Research, Tohoku University, Sendai 980-8577, Japan}

\date{\today}

\begin{abstract}
By performing time-of-flight neutron scattering measurements on a large amount of single crystals of Sr$_2$RuO$_4$, we studied detailed structure of the imaginary part of the dynamic spin susceptibility over a wide range of phase space.
In the normal state at $T=5$~K, strong incommensurate (IC) peaks were clearly observed at around $\mathbf{Q}_\text{c}=(0.3,0.3)$ up to at least $\hbar\omega=80$~meV.
In addition, our data also show strong magnetic fluctuations that exist on the ridges connecting the IC peaks around the $(\pi,\pi)$ point rather than around the $\Gamma$ point.
Our results are consistent with the semi-mean-field random phase approximation calculation for a two dimensional Fermi liquid with a characteristic energy of 5.0~meV.
Furthermore, the IC fluctuations were observed even at room temperature.
\end{abstract}

\pacs{Valid PACS appear here}
\maketitle

Despite lots of theoretical and experimental studies done since the discovery of the superconductivity below $T_\text{c}=1.5$~K~\cite{Discover,Review}, the nature of the superconductivity in ruthenate, Sr$_2$RuO$_4$, is still controversial.
The electronic structure of the ruthenate has been well studied theoretically~\cite{BandCal0,BandCal} and experimentally~\cite{ARPES}.
The $t_{2g}$ electrons of Ru$^{4+}$ ions form three bands near the Fermi surface; $d_{xz}$ and $d_{yz}$ orbitals form quasi-one-dimensional $\alpha$ and $\beta$ sheets, while $d_{xy}$ forms a two-dimensional $\gamma$ sheet~\cite{BandCal}.
Initially, it was believed that the $\alpha$ and $\beta$ sheets are responsible for the magnetic response, and the $\gamma$ sheet for the superconducting mechanism~\cite{BandTheory,gammaBand}.
The former sheets generate the incommensurate (IC) antiferromagnetic fluctuations, while the latter appears as the ferromagnetic fluctuations.

The superconducting order parameter was proposed to be a $p$-wave state with a chiral $\hat{p}_x\pm i\hat{p}_y$ symmetry~\cite{pWave,NMR}.
The origin of the $p$-wave superconductivity was thought to be the ferromagnetic fluctuations due to the $\gamma$ sheets.
However, the imaginary part of the dynamic spin susceptibility, $\chi''$, measured by inelastic neutron scattering measurement has not exhibited any detectable ferromagnetic fluctuations in both the normal and superconducting states~\cite{Neutron0,Neutron1,Neutron2,Neutron3}.
Instead, in both states gapless IC spin fluctuations were observed up to 40~meV at $\mathbf{Q}_\text{c}=(0.3,0.3)$ which come from the Fermi surface nesting of the $\alpha$ and $\beta$ sheets~\cite{Theory1}.
Furthermore, the direct experimental tests to observe the expected broken time-reversal symmetry has been conflicting.
Muon spin relaxation~\cite{muSR}, polarized neutron scattering experiment~\cite{Triplet} and Kerr effect~\cite{Kerr} measurements have been reported to support the $\hat{p}_x\pm i\hat{p}_y$ paring scenario, while scanning Hall bar and scanning superconducting quantum interference device measurements~\cite{SQUID} gave no evidence for the edge supercurrents that should be generated by such a broken time-reversal symmetry.

Very recently, theoretical work based on the weakly coupled bands reported that the quasi-one-dimensional $\alpha$ and $\beta$ sheets rather than two-dimensional $\gamma$ sheet are responsible for the triplet superconductivity and also for the lack of the edge supercurrents~\cite{1Dband}.
Since the $\alpha$ and $\beta$ sheets generate strong magnetic fluctuations, it would be crucial to obtain detailed information about dynamical spin susceptibility over a wide range of the momentum ($\mathbf{Q}$) and energy ($\hbar\omega$) phase space.
Many theoretical models~\cite{vanHove,Theory1,Theory0,Theory2,Theory3,Gamma} predict the IC spin fluctuations, but the fine structure of $\chi''$ varies in different models; while the position of the main peaks in $\chi''$ is quite stable with regard to the computational details, the exact shape of the two-dimensional ridges connecting these peaks depends very much on the exact degree of the $\alpha$/$\beta$ nesting, including the residual three-dimensional dispersion.
As a result, in some papers the ridges are largely suppressed around $\Gamma$~\cite{Theory1,Theory0,Theory2}, while in others with different parameters and an additional spin-orbit interaction they extend over the entire length of the Brillouin zone~\cite{Gamma}.
Previous neutron scattering experiments have been limited to obtain the detailed structure of $\chi''$, mainly due to small amounts of sample and also due to the triple-axis spectroscopy (TAS) that was utilized to cover only a limited area of the phase space.

Here, we report our time-of-flight (TOF) neutron scattering experiment on $\sim30$~g single crystals of Sr$_2$RuO$_4$ to show detailed structure of the spin fluctuations over a very wide range of 
$\mathbf{Q}-\hbar\omega$ phase space
 in the normal state of Sr$_2$RuO$_4$.
Our results clearly show that the strong IC peaks exist up to at least 80~meV and can be well explained by the phenomenological response function for a Fermi liquid.
Our analysis yield the characteristic energy for the spin fluctuation, $\hbar\omega_\text{SF}$ of 5.0(2)~meV; $\hbar\omega_\text{SF}$ is close to $k_\text{B}T_\text{FL}$ below which Sr$_2$RuO$_4$ exhibit two-dimensional Fermi liquid behavior in the resistivity and specific heat measurements~\cite{2DFermi,resist,SpecificHeat}.
Our data also shows that the weaker ridge scattering exists around the $(\pi,\pi)$ point rather than the $\Gamma$ point.
Furthermore, unlike in the previous neutron study~\cite{Neutron0,Neutron1}, the IC fluctuations were observed even at room temperature.

\begin{figure}[t]
\includegraphics[width=8.54cm,height=5.6cm]{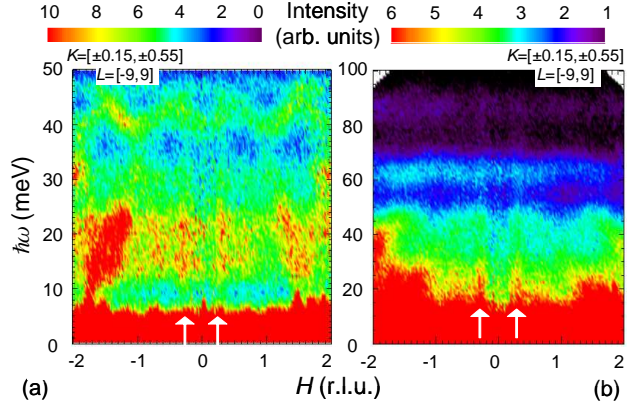}
\centering
\caption{(color online).
Contour map of the neutron scattering intensity as function of $H$ and $\hbar\omega$ at $T=5$~K.
The incident neutron energies were (a) $E_\text{i}=101.6$~meV and (b) 151.2~meV.
$S(\mathbf{Q},\hbar\omega)$ was integrated along the $K$-direction from $\pm0.15$ to $\pm0.55$ and $L$-direction from $-9$ to 9, respectively.
Arrows represent the IC peak positions.
}\label{Fig:Phonon}
\end{figure}

Five single crystals of Sr$_2$RuO$_4$ with a total mass of 29.15~g were prepared by a floating-zone method~\cite{Sample1,Sample2}.
They were co-aligned for our neutron scattering measurements performed at the TOF spectrometer 4SEASONS~\cite{4SEASONS,4SEASONS2} located at J-PARC, Tokai in Japan.
The crystals were mounted in a way that the crystallographic $c$-axis was along the incident neutron beam, which allows one to probe the scattering in the $(H\ K\ 0)$ plane for this quasi-two-dimensional system.
The crystals were put into an aluminum sample can that was then attached to a closed-cycle displex refrigerator, and the measurements were done at 5~K and 300~K.
The incident neutron energies were selected by a Fermi chopper and a combination of two disk choppers~\cite{4SEASONS2}.
Due to the unique data collection system~\cite{4SEASONS3}, for a given setup of the choppers, the scattering events from neutrons with different incident energies, $E_\text{i}$, are collected simultaneously by the detectors.
We selected two setups that collected the following sets of incident energies: one with $E_\text{i}=12.6$, 21.6, 45.5, 151.2~meV and another with $E_\text{i}=28.1$, 48.3, 101.6, 336.6~meV.
Here, we show the data obtained with three most relevant incident energies, $E_\text{i}=12.6$ (for magnetic fluctuations), and 101.6 and 151.2~meV (for the overall view of the magnetic fluctuations and lattice vibrational modes).

Figures~\ref{Fig:Phonon}(a) and \ref{Fig:Phonon}(b) show the scattering intensity with energy transfer up to $\hbar\omega=100$~meV along the $H$ direction.
There are strong bands of scattering centered at around 20~meV, 40~meV, 60~meV, and 85~meV.
They are phonons that have already been reported by the previous neutron scattering study using a thermal triple-axis spectrometer (TAS)~\cite{Phonon}.
In addition, there are strong spin fluctuations at the IC positions centered at $\mathbf{Q}_\text{c}=(0.3,0.3)$ that exist up to $\hbar\omega$ of at least 80~meV (see the white arrows), which are consistent with a previous polarized inelastic neutron scattering experiment that reported the IC peaks up to 40~meV~\cite{Neutron3}.

\begin{figure}[t]
\includegraphics[width=8.4cm,height=11.2cm]{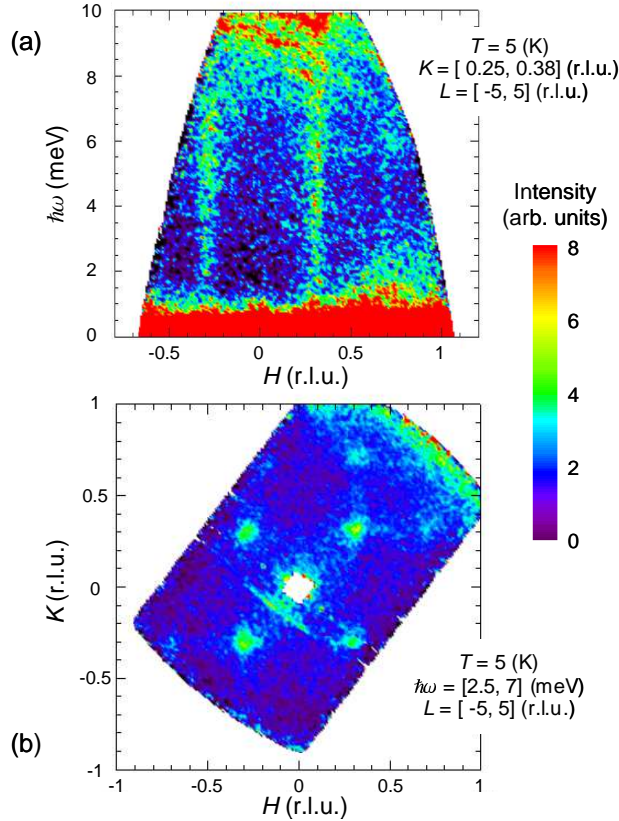}
\centering
\caption{(color online).
(a) Contour map of the neutron scattering intensity at $T=5$~K as function of $\hbar\omega$ and $H$.
The intensity was integrated over $K$ from 0.25 to 0.38 and $L$ from $-5$ to 5.
(b) Contour map at constant-$\hbar\omega$ cut the neutron scattering intensity at $T=5$~K.
The energy window was from $\hbar\omega=2.5$ to 7~meV, and the scattering was integrated over $-5\le L\le5$.
The incident neutron energy was $E_\text{i}=12.6$~meV in both panels.
}\label{Fig:Contour}
\end{figure}

The magnetic fluctuations can be more clearly seen at low energies where no strong phonon modes exist.
Figure~\ref{Fig:Contour}(a) shows the data up to $\hbar\omega=10$~meV.
The strong IC peaks are centered at the characteristic wave vector transfer, $\mathbf{Q}_\text{c}=(0.3,0.3)$.
The peak position does not change with $\hbar\omega$ up to 80~meV (see Fig.~\ref{Fig:Phonon}).
The full-width-of-the-half-maximum of the magnetic scattering along the $H$-direction does not change at low energies (see Fig.~\ref{Fig:Contour}(a)), suggesting a very high stiffness of the magnetic fluctuations.
This will be discussed in detail later.

\begin{figure}[t]
\includegraphics[width=8.4cm,height=11.2cm]{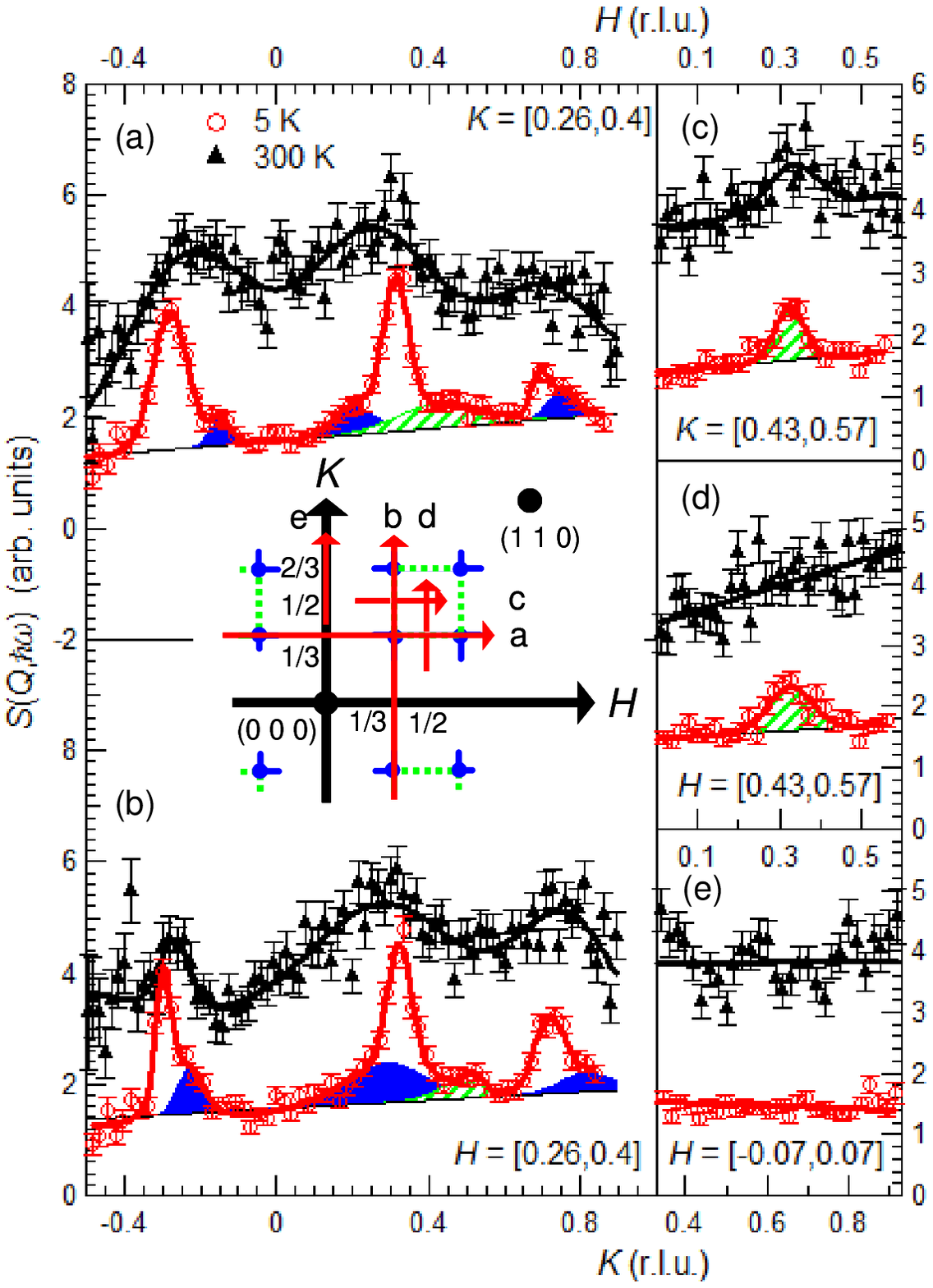}
\centering
\caption{(color online).
Several constant-$\hbar\omega$ cuts of the neutron scattering intensities at $T=5$~K (red circles) and 300~K (black triangles) along different directions in the reciprocal space as described by arrows in inset: along (a) $(H\ 0.33\ 0)$, (b) $(H\ 0.5\ 0)$, (c) $(0.33\ K\ 0)$, (d) $(0.5\ K\ 0)$, and (e) $(0\ K\ 0)$.
The energy window was from $\hbar\omega=2.5$ to 7~meV, and $L$ was integrated from $-5$ to 5.
The incident neutron energy was $E_\text{i}=12.6$~meV.
Striped areas (green) represent the ridge intensities, and solid areas (blue) represent the shoulder peaks.
Solid lines are fitting results using the combination of gaussians and the linear background.
The inset describes the schematic view of the ($H\ K\ 0$) plane in the reciprocal space.
Big solid circles (black) represent the nuclear Bragg reflections.
Small solid circles (blue) represent the IC magnetic peaks, thin lines (blue) show the positions of the shoulder peaks, and dashed lines (green) describe the ridge intensities between IC peaks~\cite{Theory1,Theory0,Neutron1}.
Five thin arrows (red) show the directions (a, b, c, d, and e) in the reciprocal plane for each constant-$\hbar\omega$ cuts shown in Fig.~\ref{Fig:Cuts}.
}\label{Fig:Cuts}
\end{figure}

In order to investigate the $\mathbf{Q}$-dependence of the low energy magnetic fluctuations, we integrated $S(\mathbf{Q},\hbar\omega)$ over $\hbar\omega$ from 2.5 to 7~meV and $L$ from $-5$ to 5 at $T=5$~K, and the result as a function of $\mathbf{Q}=(H,K)$ is plotted in Fig.~\ref{Fig:Contour}(b).
There exist several IC peaks with the characteristic wave vector of $\mathbf{Q}_\text{c}=(0.3,0.3)$.
Positions of the IC peaks are summarized in the inset of Fig.~\ref{Fig:Cuts}.
The IC magnetic fluctuations were theoretically predicted as the consequence of the Fermi surface nesting of the quasi-one-dimensional $\alpha$ and $\beta$ sheets~\cite{Theory0,Theory1} and have been observed experimentally~\cite{Neutron0,Neutron1,Neutron2,Neutron3}.
On the other hand, these peaks showed no $L$-dependence~\cite{Q2D,Neutron1,Neutron4}.

\begin{figure}[t]
\includegraphics[width=8.4cm,height=9.8cm]{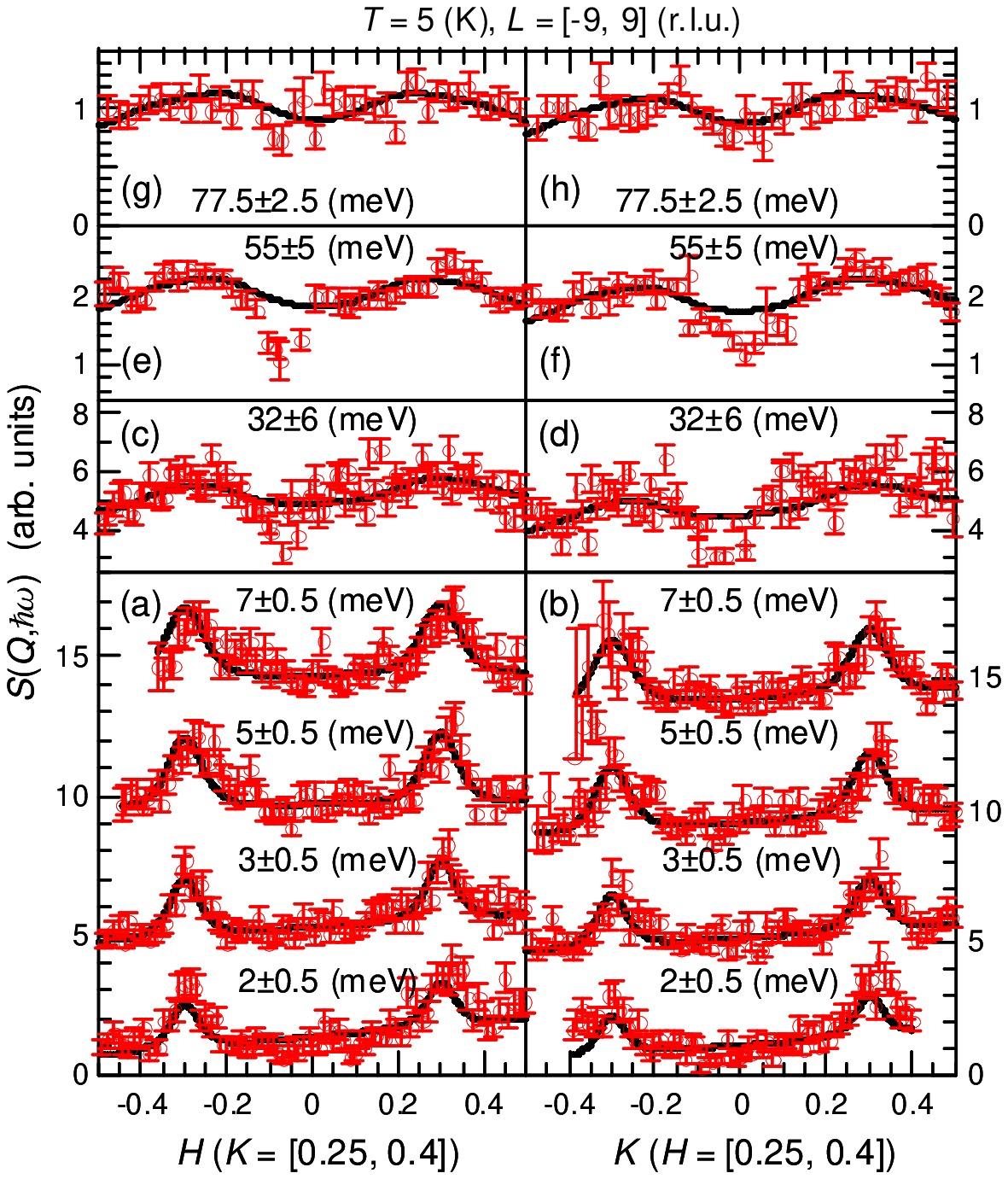}
\centering
\caption{(color online).
$H$- and $K$-dependences of the neutron scattering intensity in the normal state at (a, b) $\hbar\omega=2$, 3, 5, 7, (c, d) 32, (e, f) 55, and (g, h) 77.5~meV.
$H$ and $K$ were integrated from 0.25 to 0.4, respectively, and $L$ was integrated from $-9$ to 9.
The incident neutron energies were (a, b) $E_\text{i}=12.6$~meV, (c, d) 101.6~meV, and (e-h) 151.2~meV.
Fitting results using Eqs.~(\ref{Eq:chi}) and (\ref{Eq:SatoMaki}) with a linear background at each $\hbar\omega$ are also described by solid lines.
}\label{Fig:SatoMaki}
\end{figure}

To address the aforementioned controversy over the location of the ridge scattering, we integrated the neutron scattering intensity of the low energy excitations shown in Fig.~\ref{Fig:Contour}(b) along several different $\mathbf{Q}$-directions as shown in the inset of Fig.~\ref{Fig:Cuts}.
Figure~\ref{Fig:Cuts} shows the results.
Figures~\ref{Fig:Cuts}(a) and \ref{Fig:Cuts}(b) show the strong IC peaks at $\mathbf{Q}_\text{c}$ at 5~K (see red circles).
Using the combination of gaussians and the linear background, the peak positions are obtained as follows: $H=-0.311(1)$, 0.282(1), and 0.660(1) with $K\approx0.33$, and $K=-0.327(5)$, 0.292(2), and 0.686(5) with $H\approx0.33$.
As shown as black triangles, the IC fluctuations are broad but they survive even at 300~K.
In addition, there are some shoulder peaks shown by the blue gaussians at $H=-0.179(1)$, 0.175(3), and 0.720(4) with $K\approx0.33$, and at $K=-0.260(2)$, 0.283(1), and 0.781(9) with $H\approx0.33$.
These two features have already been observed in the previous neutron scattering study~\cite{Neutron1}.
For the possible ridge scattering, we plot the data across the ridges around $(\pi,\pi)$ (see the inset for the directions, and Figs.~\ref{Fig:Cuts}(c) and \ref{Fig:Cuts}(d) for the data) and around the $\Gamma$ point (Fig.~\ref{Fig:Cuts}(e)).
In contrast to the previous study~\cite{Neutron1}, our TOF data show strong ridge scattering around $(\pi,\pi)$ (at $H=0.326(1)$ with $K\approx0.5$ and $K=0.327(1)$ with $H\approx0.5$ in Figs.~\ref{Fig:Cuts}(c) and \ref{Fig:Cuts}(d), respectively) rather than around the $\Gamma$ point (Fig.~\ref{Fig:Cuts}(e)).
This can be seen in Figs.~\ref{Fig:Cuts}(a) and \ref{Fig:Cuts}(b) as well; there are intensities between the peaks at $H\ (\text{or}\ $K$)\ =0.3$ and 0.7 represented by green gaussian, but no additional intensity to the background between $H\ (\text{or}\ $K$)\ =-0.3$ and 0.3 except for the shoulder intensities.
It is an unresolved question how the previous TAS experiment has yielded the different ridge scattering~\cite{Neutron1}.

To study $\hbar\omega$-dependence of the IC magnetic fluctuations at 5~K, we obtained the constant-$\hbar\omega$ cuts at several different energy transfers as a function of $H$ or $K$.
Figure~\ref{Fig:SatoMaki} shows the results.
Figures~\ref{Fig:SatoMaki}(a) and \ref{Fig:SatoMaki}(b) show the low energy IC peaks can be easily extracted from the data shown in Fig.~2(a).
Figures~\ref{Fig:SatoMaki}(c-h) show the IC peaks at high energies where phonons are weak and the magnetic scattering can be extracted from the data shown in Figs.~1(a) and (b).
Our data clearly show that IC magnetic fluctuations in Sr$_2$RuO$_4$ survive all the way up to $\sim80$~meV.
The magnetic scattering intensity is related to the imaginary part of the dynamic susceptibility, $\chi''(\mathbf{Q},\hbar\omega)$, using the fluctuation dissipation theorem,
\begin{equation}\label{Eq:chi}
S(\mathbf{Q},\hbar\omega)=\left|F(Q)\right|^2\frac{\chi''(\mathbf{Q},\hbar\omega)}{1-\text{exp}\left(-\hbar\omega/k_\text{B}T\right)}.
\end{equation}
The magnetic form factor of Sr$_2$RuO$_4$, $F(Q)$, was obtained from reference~\cite{Neutron4}.
For a quantitative analysis, we fit the data to the general form of the phenomenological response function used to describe a Fermi liquid system~\cite{Hayden}; 
\begin{equation}\label{Eq:SatoMaki}
\chi''(\mathbf{Q},\hbar\omega)=\sum\limits_{\mathbf{Q}_\text{c}}\frac{\chi_\delta\kappa_0^4\left(\hbar\omega/\hbar\omega_\text{SF}\right)}{\left[\kappa_0^2+\left(\mathbf{Q}-\mathbf{Q}_\text{c}\right)^2\right]^2+\left(\hbar\omega/\hbar\omega_\text{SF}\right)^2\kappa_0^4}
\end{equation}
where $\chi_\delta$, $\kappa_0$, $\hbar\omega_\text{SF}$, and $\mathbf{Q}_\text{c}$ are parameters for the peak intensity, the sharpness of the peak, the characteristic energy of the spin fluctuations, and the IC peak position~\cite{SatoMaki1,SatoMaki2}, which are all independent of $\hbar\omega$.
We fitted the $H$- and $K$-dependences at each $\hbar\omega$ positions simultaneously to Eqs.~(\ref{Eq:chi}) and (\ref{Eq:SatoMaki}) with a linear background.
$\mathbf{Q}_\text{c}$ was obtained to be $\mathbf{Q}_\text{c}=(0.301(3),0.303(4))$ from fitting the 3~meV data (Figs.~\ref{Fig:SatoMaki}(a) and \ref{Fig:SatoMaki}(b)), and fixed for all other energies.
As shown by the solid lines in Fig.~\ref{Fig:SatoMaki}, Eq.~(\ref{Eq:SatoMaki}) describes our data well over the entire energy range.
The optimum parameters obtained from the best fit are $\chi_\delta=10.6(3)$, $\kappa_0=0.053(2)$~r.l.u. ($=0.086(4)$~$\text{\AA}$$^{-1}$) and $\hbar\omega_\text{SF}=5.0(2)$~meV.
The fact that $\hbar\omega_\text{SF}/k_\text{B}=58(3)$~K is close to $T_\text{FL}$ may indicate that the two-dimensional IC spin fluctuations may be responsible for the Fermi liquid behavior.

Our inelastic neutron scattering measurements on Sr$_2$RuO$_4$ over a very wide range of the $\mathbf{Q}-\hbar\omega$ phase space revealed the three components of the magnetic fluctuations; the strong IC spin fluctuations at 5~K centered at $\mathbf{Q}_\text{c}=(0.301(3),0.303(4))$ that extend up to at least 80~meV with the characteristic energy of $\hbar\omega_\text{SF}=5.0(2)$~meV, and the weaker so-called shoulder scattering and the ridge scattering at low energy transfer.
Our data clearly shows that the ridge scattering is strong around the $(\pi,\pi)$ rather than around the $\Gamma$ points.
We have also shown that the IC fluctuations survive up to 300~K that is much higher than $T_\text{c}$.
Together with the previous observation that the magnetic fluctuations do not change when the system enters the superconducting phase below $T_\text{c}$~\cite{Neutron1}, our results suggest that the magnetic and superconducting properties of Sr$_2$RuO$_4$ are not connected.

We thank Y.-B. Kim and I. Mazin for helpful discussion, and Y. Qiu for his help with analyzing the data.
Works at the University of Virginia were supported by the US NSF under Agreement No.~DMR-0903977.


\begin{thebibliography}{60}

\bibitem{Discover}
Y. Maeno, \textit{et} \textit{al.}, 
Nature \textbf{372}, 532 (1994).

\bibitem{Review}
A. P. Mackenzie and Y. Maeno, 
Rev. Mod. Phys. \textbf{75}, 657 (2003).

\bibitem{BandCal0}
T. Oguchi, 
Phys. Rev. B \textbf{51}, 1385 (1995).

\bibitem{BandCal}
I. I. Mazin \textit{et} \textit{al.}, 
Phys. Rev. Lett. \textbf{79}, 733 (1997).

\bibitem{ARPES}
A. Damascelli \textit{et} \textit{al.}, 
Phys. Rev. Lett. \textbf{85}, 5194 (2000).

\bibitem{gammaBand}
T. Nomura \textit{et} \textit{al.}, 
J. Phys. Soc. Jpn. \textbf{71}, 1993 (2002).

\bibitem{BandTheory}
Y. Yoshioka \textit{et} \textit{al.}, 
J. Phys. Soc. Jpn. \textbf{78}, 074701 (2009).

\bibitem{pWave}
T. M. Rice and M. Sigrist, 
J. Phys. Condens. Matter \textbf{7}, L643 (1995).

\bibitem{NMR}
K. Ishida \textit{et} \textit{al.}, 
Nature \textbf{396}, 658 (1998).

\bibitem{Neutron0}
Y. Sidis \textit{et} \textit{al.}, 
Phys. Rev. Lett. \textbf{83}, 3320 (1999).

\bibitem{Neutron1}
M. Braden \textit{et} \textit{al.}, 
Phys. Rev. B \textbf{66}, 064522 (2002).

\bibitem{Neutron2}
F. Servant \textit{et} \textit{al.}, 
Phys. Rev. B \textbf{65}, 184511 (2002).

\bibitem{Neutron3}
M. Braden \textit{et} \textit{al.}, 
Phys. Rev. Lett. \textbf{92}, 097402 (2004).

\bibitem{Theory1}
I. I. Mazin and D. J. Singh, 
Phys. Rev. Lett. \textbf{82}, 4324 (1999).

\bibitem{muSR}
G. M. Luke \textit{et} \textit{al.}, 
Nature \textbf{394}, 558 (1998).

\bibitem{Triplet}
J. A. Duffy \textit{et} \textit{al.}, 
Phys. Rev. Lett. \textbf{85}, 5412 (2000).

\bibitem{Kerr}
J. Xia \textit{et} \textit{al.}, 
Phys. Rev. Lett. \textbf{97}, 167002 (2006).

\bibitem{SQUID}
J. R. Kirtley \textit{et} \textit{al.}, 
Phys. Rev. B \textbf{76}, 014526 (2007).

\bibitem{1Dband}
S. Raghu \textit{et} \textit{al.}, 
Phys. Rev. Lett. \textbf{105}, 136401 (2010).

\bibitem{vanHove}
T. Nomura \textit{et} \textit{al.}, 
J. Phys. Soc. Jpn. \textbf{69}, 1856 (2000).

\bibitem{Theory0}
I. Eremin \textit{et} \textit{al.}, 
Europhys. Lett. \textbf{58}, 871 (2002).

\bibitem{Theory2}
D. K. Morr \textit{et} \textit{al.}, 
Phys. Rev. Lett. \textbf{86}, 5978 (2001).

\bibitem{Theory3}
T. Takimoto, 
Phys. Rev. B \textbf{62}, R14641 (2000).

\bibitem{Gamma}
K.-K. Ng \textit{et} \textit{al.}, 
J. Phys. Soc. Jpn. \textbf{69}, 3764 (2000).

\bibitem{2DFermi}
Y. Maeno \textit{et} \textit{al.}, 
J. Phys. Soc. Jpn. \textbf{66}, 1405 (1997).

\bibitem{resist}
N. E. Hussey \textit{et} \textit{al.}, 
Phys. Rev. B \textbf{57}, 5505 (1998).

\bibitem{SpecificHeat}
A. P. Mackenzie \textit{et} \textit{al.}, 
J. Phys. Soc. Jpn. \textbf{67}, 385 (1998).

\bibitem{Sample1}
S. I. Ikeda \textit{et} \textit{al.}, 
J. Crystal Growth \textbf{237-239}, 787 (2002).

\bibitem{Sample2}
Z. Q. Mao \textit{et} \textit{al.}, 
Mater. Res. Bull. \textbf{35}, 1813 (2000).

\bibitem{4SEASONS}
R. Kajimoto \textit{et} \textit{al.}, 
J. Neutron Research \textbf{15}, 5 (2007).

\bibitem{4SEASONS2}
R. Kajimoto \textit{et} \textit{al.}, 
Nucl. Instr. and Meth. A \textbf{600}, 185 (2009).

\bibitem{4SEASONS3}
M. Nakamura \textit{et} \textit{al.}, 
J. Phys. Soc. Jpn. \textbf{78}, 093002 (2009).

\bibitem{Phonon}
M. Braden \textit{et} \textit{al.}, 
Phys. Rev. B \textbf{76}, 014505 (2007).

\bibitem{Q2D}
F. Servant \textit{et} \textit{al.}, 
Solid State Commun. \textbf{116}, 489 (2000).

\bibitem{Neutron4}
T. Nagata \textit{et} \textit{al.}, 
Phys. Rev. B \textbf{69}, 174501 (2004).

\bibitem{Hayden}
S. M. Hayden \textit{et} \textit{al.}, 
Phys. Rev. Lett. \textbf{84}, 999 (2000).

\bibitem{SatoMaki1}
A. J. Millis \textit{et} \textit{al.}, 
Phys. Rev. B \textbf{42}, 167 (1990).

\bibitem{SatoMaki2}
Y. Zha \textit{et} \textit{al.}, 
Phys. Rev. B \textbf{54}, 7561 (1996).

\end{thebibliography}
\end{document}